\documentclass[twocolumn,prl]{revtex4}

\usepackage{graphicx}
\begin{document}

\preprint{Draft-Jan. 23 '04}

\title{Softening of Cu-O bond stretching phonon in tetragonal HgBa$_2$CuO$_{4+\delta}$}

\author{H. Uchiyama$^1$, A.Q.R. Baron$^2$, S. Tsutsui$^2$, Y. Tanaka$^3$, W.-Z. Hu$^1$, A.Yamamoto$^1$,  S. Tajima$^1$, and  Y. Endoh$^4$}
\affiliation{
$^{1}$Superconductivity Research Laboratory, International Superconductivity
Technology Center, 1-10-13 Shinonome, Koto-ku, Tokyo 135-0062, Japan.\\
2 SPring-8/JASRI,
 Mikazuki Hyogo, 679-5198, Japan.\\
3 SPring-8/RIKEN,
 Mikazuki, Hyogo, 679-5148, Japan.\\
4 Institute for Material Research, Tohoku University, Sendai, 980-8577, Japan.
}%

\date{\today}
\begin{abstract}
Phonons in nearly optimally doped HgBa$_2$CuO$_{4+\delta}$  were studied by  inelastic X-ray scattering. The dispersion of the low energy modes is well described by a shell model, while the Cu-O bond stretching mode at high energy shows strong softening towards the zone boundary, which deviates strongly from the model. This seems to be common in the hole-doped high-$T_\mathrm{c}$ superconducting cuprates, and, based on this work, not related to a lattice distortion specific to each material. \end{abstract}

\maketitle

   During the history of high-$T_\mathrm{c}$  superconductivity research, there have been many pieces of evidence pointing to strong electron-phonon coupling in these system.  For example, electronic changes such as the superconducting transition or  pseudo-gap opening definitely induce phonon renormalization effects\cite{IntroRaman}. Nevertheless, the phonon contribution to the electronic properties has been unclear or considered  unimportant.  Even though the Fermi surface shape favors formation of a charge-density-wave (CDW), no CDW gap has been observed. Instead, the parent compound is a Mott insulator.  The carrier scattering mechanism responsible for the $T$-linear in-plane resistivity at the optimal doping is considered to be dominated by spin-fluctuation rather than by phonons\cite{resistivity}.
   
   Recently the role of phonons in the electronic state has been reassessed.  The isotope effect on the magnetic penetration depth suggests a tight connection of the oxygen vibrations to the superfluid density or the effective mass relevant to the paring interaction\cite{Isotope}.  Photoemission suggests that electron-phonon interaction causes a kink at 50-80 meV in the electronic dispersion for many cuprates\cite{Lanzara}.
   
While phonon dispersion in high-$T_\mathrm{c}$ superconducting cuprates (HTSC) has been studied for a long time\cite{Pintschovius}, recent interest has been stimulated by inelastic neutron scattering (INS)\cite{Mook,PintschoviusLa214,PintschoviusYBCO,ElPhLa214,ElPhYBCO}  and inelastic X-ray scattering (IXS)\cite{Nd214IXS} studies that show anomalies in the copper-oxygen phonon branches.  These anomalies were discussed in terms of spatial charge fluctuation\cite{PintschoviusYBCO,PintschoviusLa214,Mook} and connection to the mechanism of superconductivity\cite{ElPhLa214,ElPhYBCO}.    However, the studies to date have been limited to  YBa$_2$Cu$_3$O$_{7-\delta}$ (YBCO)  and  La$_{2-x}$Sr$_x$CuO$_4$ (LSCO), except for the recent work with electron doped Nd$_{2-x}$Ce$_x$CuO$_4$ (NCCO).  Thus, in order to really understand the physics of these phonon anomalies,  it is strongly desirable to study phonons of different HTSCs.

HgBa$_2$CuO$_{4+\delta}$ (Hg1201) is, in principle, an excellent candidate for phonon studies.  The crystal structure is tetragonal, forming a completely square and flat CuO$_2$ plane, as compared to the complexity of the chain structure in YBCO and the tetragonal to orthorhombic phase transitions in LSCO\cite{La214lattice}.  Moreover, the absence of a 1/8-anomaly suggests that static stripes play a small role in Hg1201\cite{Yamamoto}.  Hg1201 is also interesting because the structure is similar to LSCO while $T_\mathrm{c}$ ($\sim$98 K) is similar to YBCO, and, indeed, members of this (Hg based) family of HTSC also have the highest $T_\mathrm{c}$ demonstrated to date ($T_\mathrm{c}$ = $\sim$135 K).  Despite these favorable properties, the Cu-O phonons in this material have not been studied, in large part because the material is not amenable to neutron scattering  due to the very small size of the crystals.  

We have performed IXS measurements on nearly optimally doped Hg1201, mapping the dispersion of the $\Delta_1$ symmetry phonons, including the Cu-O bond stretching mode.  It is the small beam size in IXS experiments that makes it even conceivable to measure samples of size $<$0.1 mm$^3$.  However, even so, this is a hard experiment since the sample contains heavy elements (Hg and Ba) which reduce the illuminated sample volume by a factor of about 3 compared to  previously studied NCCO by IXS\cite{Nd214IXS}.  Thus, in a recent report of IXS from this material\cite{Hg1201IXS} the relevant mode was only barely visible at one momentum transfer and its dispersion was not measured.  However, at BL35XU of SPring-8\cite{Baron} we obtained high quality data, allowing us to show that this mode softens similarly to optimally doped YBCO and LSCO\cite{PintschoviusYBCO,PintschoviusLa214}.  Thus we can conclude the softening is a common feature in the hole-doped HTSCs, independent of lattice distortion and transition temperature.

\begin{figure}
\includegraphics[width=1\linewidth,clip=]{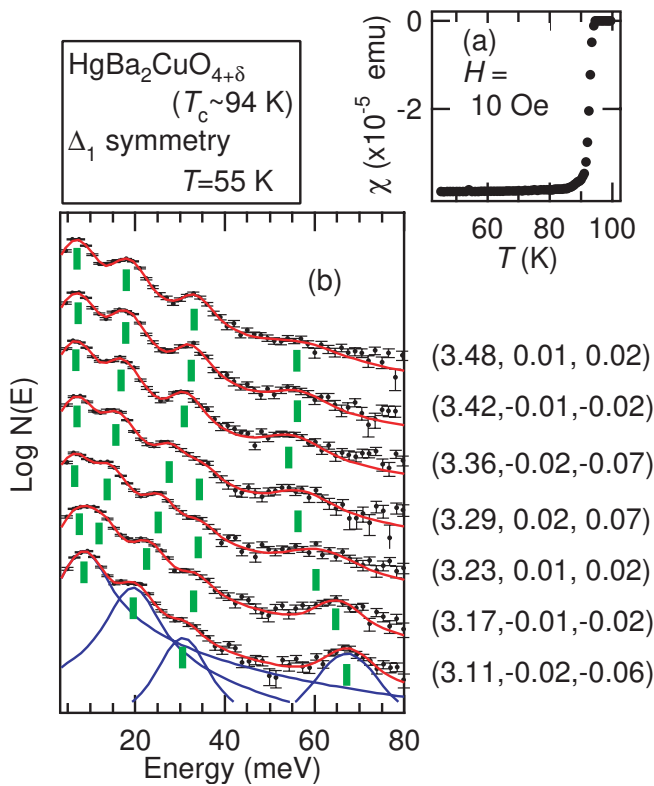}
\caption{\label{fig1}(a) Magnetic
susceptibility  of the studied crystal HgBa$_2$CuO$_{4+\delta}$. (b) IXS data  taken along the direction close to [$\xi$00] with experimental error bars (black), and the fitting curves (red)  with several peaks (blue curves as an example for $\xi$=3.11). The phonon branches (peak positions) are indicated by the vertical bars (green).}
\end{figure}
 
Single crystals of Hg1201 were prepared with a solid-state reaction method\cite{Hu},
from  HgO (99.9 \%), CuO (99.999 \%)  and BaO purified from BaCO$_3$ (99.995 \%).   The crystal with the size of 0.3$\times$0.3$\times$0.2 mm$^{3}$ was annealed at 340
$^\circ$C in O$_2$ for 1 week. Magnetic susceptibility measurement shows a sharp superconductivity transition at 94 K, indicating that the sample is nearly optimally doped (see  Fig.\ 1(a)). The crystallinity was
examined by X-ray diffraction and there was no trace of detectable impurity phase.   The IXS spectrometer\cite{Baron} employed backscattering at the silicon (888) reflection, providing $\sim$3$\times$10$^{10}$  photons/s (spot size $\sim$150$\times 100$ $ \mu m ^2$) onto the sample in a 4 meV bandwidth at 15.816 keV.   Four spherically curved analyzer crystals (each with an independent detector) were used on the 10 m horizontal arm to analyze the scattered radiation.Ê The energy resolution (6.0 - 6.3 meV, depending on analyzer crystal) was determined by measuring the elastic scattering of a standard poly-methyl methacrylate (PMMA) sample near to its structure factor maximum at 1 \AA$ ^{-1}$, while the momentum resolution was set by the full (95 mm) diameter of the crystals, maximizing count rate.Ê The sample  was mounted in a closed cycle He cryostat, with measurements performed at 55 K (well below $T_\mathrm{c}$=94 K).  Typically, about 20 hours was needed to collect sufficient statistics in one spectrum (e.g. data in Fig.\ 1(b)).

IXS spectra were measured  along the direction close to [$\xi$00], where $\xi$ is varied from 3.11 to 3.48. The simultaneous use of 4 analyzers  gives actual momenta slight deviation from ($\xi$00) (exact values given in  Fig.\ 1(b)). In this arrangement, the
$\Delta _1$-symmetry phonon modes, which includes 8 branches in
Hg1201, should have strong scattering cross section. Among the eight, one is the longitudinal
acoustic (LA) mode, two are the $c$-polarized modes of apical oxygen and barium, which are
Raman active with $A_{1g}$-symmetry  at $\Gamma$. The others are the $a$-polarized phonons that are $E_u$ longitudinal optical (LO) modes at $\Gamma$. Fig.\ 1(b) shows the phonon distribution curves taken along the direction close to [$\xi$00] with experimental error bars. At each momentum, an elastic peak  obtained from  the standard material was subtracted. 
   The peak positions indicated by short bars are determined by a voigt
fit function, and the results of peak fitting are shown by red lines.  Full components for fitting are indicated only for  $\xi$=3.11 (blue lines).  Some of the 8 phonon branches are
not detected, because of their weak  intensity. Peaks at energies  lower than
50 meV have  FWHM   $\sim$7 meV, similar to that of the elastic peak. The highest energy peak over 50 meV has a wider FWHM, $>$10 meV. All presented data were collected at
55K (some spectra at 110 K showed negligible changes relative to 55K and are not shown).

\begin{figure}
\includegraphics[width=1\linewidth,clip=]{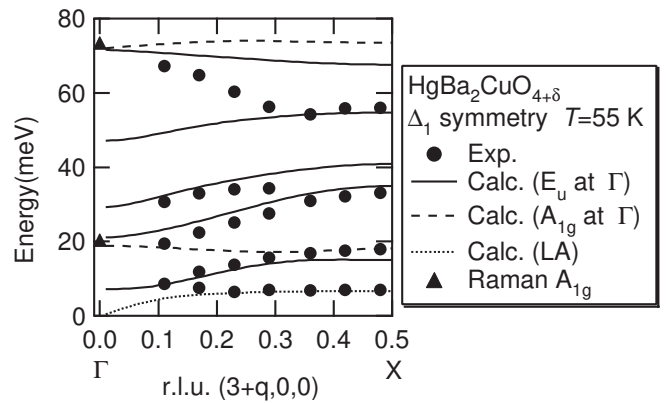}
\caption{\label{fig2} Comparison of the experimental data (circles) with the calculation (lines).    The Raman data in ref.[19] are indicated by  triangles. }
\end{figure}

 The phonon dispersion (from our fits) is shown in Fig.\ 2 along with a simple shell model
similar to that used previously to simulate cuprate phonon dispersion\cite{PintschoviusYBCO,Shell,Nd214IXS}. The parameters for this
model were based on a common set for the cuprates\cite{Shell}, with slight modification for Hg1201 \cite{Hg1201shell}, and
to fit the Raman $A_{1g}$ data\cite{Raman}, and (though independently determined) were similar to those used in ref.\ \cite{Hg1201IXS}. Calculations were done using the ÒOpenphononÓ code\cite{Mirone} with carrier screening included by a Thomas-Fermi potential, $e^{-\kappa _sr}/ r$ ( $\kappa _s$ $\sim$0.41 \AA$ ^{-1}$). For low energies, this simple shell model
provides a reasonable description of the phonon dispersion, as has also been the case for other
cuprates including LSCO, YBCO, and NCCO\cite{Shell,Pintschovius,Nd214IXS}. However, in the region above 50 meV the
agreement is not so good, in particular for the highest energy $c$-polarized apical oxygen mode
and the next highest $a$-polarized Cu-O bond stretching mode, and the Cu-O bond bending mode.
Similar problems have been noted in calculations of YBCO\cite{PintschoviusYBCO}, NCCO\cite{Nd214IXS}, and LSCO\cite{PintschoviusLa214}.
 
 Modification of the shell model by including a next-nearest neighbor oxygen interaction, similar
to ref.\ \cite{LNOTranq}, was attempted. Such an interaction phenominalogically explains a Jahn-Teller type
electron-phonon coupling that results in a rhombic distortion of the CuO$_2$ square\cite{Weber}. This
interaction changes only the high-energy mode dispersion. Fig.\ 3(b) shows the results from this
modified shell model: the apical oxygen mode softens slightly while the bond stretching mode
softens more - in fact the strength of the coupling was chosen to match the experimental energy
at $q$=3.48. However, this still does not provide an adequate model of the high energy phonon
dispersion, failing to reproduce the sharp minimum in dispersion at $q$$\sim$0.3.

\begin{figure}
\includegraphics[width=1\linewidth,clip=]{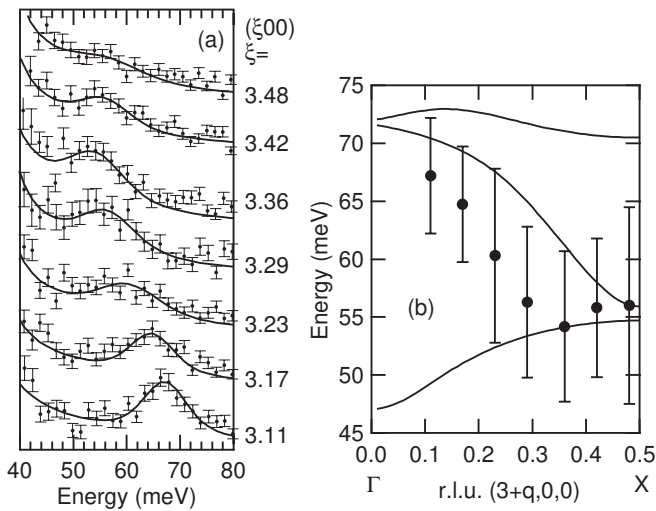}
\caption{\label{fig3}  (a) Enlarged spectra taken close to the
bond-stretching mode in a linear scale(see also Fig.\ 1(b)).
(b) The lines show the shell model calculation in which the interaction between the next nearest neighbor oxygens in the CuO$_2$-plane is added.   The lines indicate (top to bottom)   the $c$-polarized apical oxygen mode,  the $a$-polarized Cu-O bond stretching mode, and the  $a$-polarized in-plane Cu-O bending mode, respectively.  The vertical bars indicate the FWHM of the peaks  determined in  fitting in Fig.\ 1(b).}
\end{figure}

 The assignment of the measured dispersion to the bond stretching mode deserves some
additional consideration, given the presence, in calculation, of the other two modes in this energy
region. The grounds for this assignment are the intensities and broadened FWHM of the
observed peaks (shown by the bars in Fig.\ 3(b)). As regards intensity, our modified shell model
shows that apical oxygen mode intensity is so weak as to be unobservable in our experiment,
and, while stronger than the apical oxygen mode, the bond bending mode is also much weaker
than the bond-stretching mode. Furthermore, the calculation also shows that the bond stretching
mode, like the data, falls of with intensity as the zone boundary is approached ($q$=0.5), similar to
the data (a structure factor effect). As to the FWHM, it was reported for YBCO and LSCO that the bond bending mode shows a narrow FWHM close to the resolution limit\cite{PintschoviusYBCO2, privatecomm}, while the broad FWHM was observed for the stretching mode in LSCO\cite{PintschoviusLa214}. The broad FWHM plotted in Fig.\ 3 is thus another support for the bond stretching mode. Since the observed peak approaches the calculated branch for the bending mode near the zone boundary, we cannot completely exclude a possibility of mixing of the the bending modes above $q$$\sim$0.3. However, judging from the broad FWHM, we expect a dominant contribution of the bond stretching mode also to the peaks at  $q$$>$0.3.

\begin{figure}
\includegraphics[width=1\linewidth,clip=]{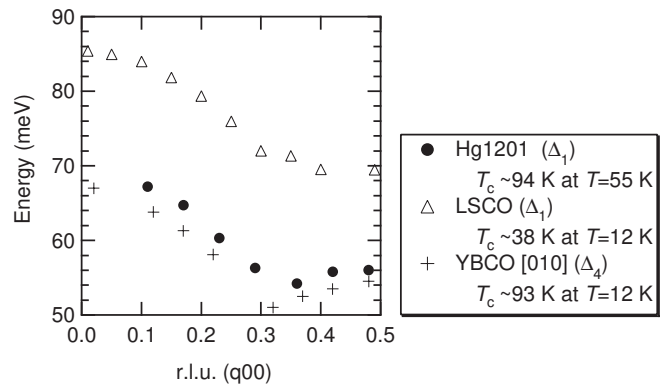}
\caption{\label{fig4} Comparison of peak position in hole-doped cuprates (YBCO[6], LSCO[9] and Hg1201).
For the YBCO, the equivalent $\Delta_4$ symmetry modes are shown for simplicity[6].
}
\end{figure}

The softening of the  Cu-O  stretching mode is also observed in LSCO\cite{Pintschovius, PintschoviusLa214,ElPhLa214} and YBCO\cite{Pintschovius, PintschoviusYBCO,ElPhYBCO}. Here, we focus on the hole-doped HTSC (LSCO, YBCO, Hg1201), as the limited data available for the electron doped materials\cite{Nd214IXS} apparently shows a different behavior\cite{comment}. Figure 4 shows a comparison of the bond stretching branches for Hg1201 (present results), LSCO\cite{PintschoviusLa214} and YBCO\cite{PintschoviusYBCO}.  All three materials show similar softening of the Cu-O bond stretching mode, irrespective of the difference in crystal structure, band dispersions and $T_\mathrm{c}$. This implies that the similar type of  electron-phonon coupling affects the CuO$_2$ plane in the hole-doped HTSC. 

It should be noted that the bond stretching phonon softening in the [100] direction is also observed in doped La$_2$NiO$_4$\cite{LNOTranq,LNOPintschovius}, LaMnO$_3$\cite{BradenLMO, ZhangLMO}, and BaBiO$_3$\cite{Braden}. Therefore, the softening of the bond stretching phonon  is a common property  in  the perovskite related materials. However, the cuprates show softening only  in the [100]  direction while for the other materials    the 
softening is observed  both in the [100]  and [110] directions.  Although the softening along the [110] direction is possibly induced by the Fermi surface nesting\cite{Weber2}, the softening in the [100] direction can not be understood in the nesting scenario.
  
Various theoretical models have been used to explain the softening of the bond stretching phonon in the cuprates\cite{Weber, Krakauer, Bohnen,Falter,Bishop, Mihailovic,Kaneshita, Park,Bishop2,Tachiki}. One is a shell model, as demonstrated in Fig.\ 3(b), while another  is to treat band effects explicitly in a calculation such as LAPW (linearized augmented plane-wave linear-response) method.  This model   gives a gradual softening of the bond stretching phonon towards the zone boundary\cite{ Krakauer, Bohnen}.  However, a rapid drop of the phonon branch at an intermediate $q$  is not explained by this model.

   The abrupt phonon softening we observe suggests introduction of some kind of charge modulation\cite{Bishop, Mihailovic,Bishop2,Kaneshita, Park,Falter, Tachiki}.  The mechanisms of charge inhomogeneity are different in different models.  For example, the spin stripe order accompanied by the charge stripes is considered as a source of the additional feature in refs.\ \cite{Kaneshita, Park}, while the Jahn-Teller instability is a driving force in refs.\ \cite{Bishop, Mihailovic}.  In order to clarify the mechanism,  further studies of various materials,  and  comparison between theory and experiment, is  needed.

In summary, we have studied the phonon dispersion of Hg1201 by IXS. While the low-energy branches are well described by  the poorly screened shell model, the Cu-O bond stretching branch is found  to be radically softened towards $q$$\sim$0.3 along the [100] direction. 
This softening is quite similar to  LSCO and YBCO, and thus we consider it a common  property of  the hole-doped HTSC, and independent of lattice distortion such as the orthorhombicity and the buckling of CuO$_2$ plane.  The  weak minimum with  $q$  hints at some kind of inhomogeneous charge distribution.  

We are grateful to L. Pintschovius  for useful suggestions.
This  work was supported by the New Energy and Industrial
Technology Department Organization (NEDO) as Collaborative
Research and Development of Fundamental Technologies for
Superconductivity Applications. The synchrotron radiation experiments were performed at the BL35XU in the SPring-8 with the approval of the Japan Synchrotron Radiation Research Institute (JASRI) (Proposal No.2002B0287-ND3-np, 2003A0555-ND3-np). H.U. acknowledges support by the JSPS Research Fellowships for Young Scientists. S.T. was financially supported by a Grant-in-Aid for Young Scientists 
from the Japan Society for the Promotion of Science.  

\end{document}